\documentclass[8.5pt,twoside,twocolumn]{article}
\usepackage[super,sort&compress,comma]{natbib} 
\usepackage{mhchem}
\usepackage{amssymb}
\usepackage{amsmath}
\usepackage{color}% color
\usepackage{sectsty}
\usepackage{balance} 

\usepackage{graphicx} %eps figures can be used instead
\usepackage{lastpage}
\usepackage[format=plain,justification=raggedright,singlelinecheck=false,font=small,labelfont=bf,labelsep=space]{caption} 
\usepackage{fancyhdr}
\begin{document}

\thispagestyle{plain}
\renewcommand{\headrulewidth}{1pt}
\renewcommand{\thefootnote}{\fnsymbol{footnote}}
\renewcommand\footnoterule{\vspace*{1pt}% 
\hrule width 3.4in height 0.4pt \vspace*{5pt}} 
\setcounter{secnumdepth}{5}

\makeatletter 
\def\subsubsection{\@startsection{subsubsection}{3}{10pt}{-1.25ex plus -1ex minus -.1ex}{0ex plus 0ex}{\normalsize\bf}} 
\def\paragraph{\@startsection{paragraph}{4}{10pt}{-1.25ex plus -1ex minus -.1ex}{0ex plus 0ex}{\normalsize\textit}} 
\renewcommand\@biblabel[1]{#1}            
\renewcommand\@makefntext[1]% 
{\noindent\makebox[0pt][r]{\@thefnmark\,}#1}
\makeatother 
\renewcommand{\figurename}{\small{Fig.}~}
\sectionfont{\large}
\subsectionfont{\normalsize} 

\renewcommand{\headrulewidth}{1pt} 
\renewcommand{\footrulewidth}{1pt}
\setlength{\arrayrulewidth}{1pt}
\setlength{\columnsep}{6.5mm}
\setlength\bibsep{1pt}

\twocolumn[
  \begin{@twocolumnfalse}
\noindent\LARGE{\textbf{Bulk modulus of the nanoparticle system in concentrated magnetic fluids and local field-induced structural anisotropy}}
\vspace{0.6cm}

\noindent\large{\textbf{E.~Wandersman,\textit{$^{a,b}$} A.~C\={e}bers,\textit{$^{c}$} E.~Dubois,\textit{$^{a}$} G.~M\'eriguet,\textit{$^{a}$} A.~Robert,\textit{$^{d, e}$} and R.Perzynski,\textit{$^{\ast}$$^{a}$}}}\vspace{0.5cm}
%Please note that \ast indicates the corresponding author(s) but no footnote text is required. 

\vspace{0.6cm}
%Please do not change this text.

\noindent \normalsize{In the present study we probe the bulk modulus and the structure of concentrated magnetic fluids by Small Angle X-ray Scattering. The electrostatically stabilized nanoparticles experience a repulsive interparticle potential modulated by dipolar magnetic interactions. On the interparticle distance lengthscale, we show that nanoparticles are trapped under-field in oblate cages formed by their first neighbours. We propose a theoretical model of magnetostriction for the field-induced deformation of the cage. This model captures the anisotropic features of the experimentally observed scattering pattern on the local scale in these strongly interacting colloidal dispersions.}
\vspace{0.5cm}
 \end{@twocolumnfalse}
  ]

\section{Introduction}

%Footnotes
%\footnotetext{\dag~Electronic Supplementary Information (ESI) available: [details of any supplementary information available should be included here]. See DOI: 10.1039/b000000x/}

%Please use \dag to cite the ESI in the main text of the article.
%If you article does not have ESI please remove the the \dag symbol from the title and the above footnotetext.
\footnotetext{\textit{$^{a}$~UPMC - Lab. PECSA - UMR 7195 CNRS-UPMC-ESPCI, case 51, 4 place Jussieu, F-75005 Paris, France. E-mail: regine.perzynski@upmc.fr}}
\footnotetext{\textit{$^{b}$~UPMC - Lab. Jean Perrin - FRE 3231 CNRS-UPMC, case 114, 4 place Jussieu, F-75005 Paris, France. }}
\footnotetext{\textit{$^{c}$~University of Latvia, Zellu-8, Riga, LV-1002, Latvia.}}
\footnotetext{\textit{$^{d}$~European Synchrotron Radiation Facility - 6 rue Jules Horowitz BP 220, F-38043 Grenoble Cedex 9, France.}}
\footnotetext{\textit{$^{e}$~SLAC National Accelerator Laboratory, Linac Coherent Light Source, 2575 Sand Hill Rd, Menlo Park CA 94025 - US. }}

%additional addresses can be cited as above using the lower-case letters, c, d, e... If all authors are from the same address, no letter is required

%\footnotetext{\ddag~Additional footnotes to the title and authors can be included \emph{e.g.}\ `Present address:' or `These authors contributed equally to this work' as above using the symbols: \ddag, \textsection, and \P. Please place the appropriate symbol next to the author's name and include a \texttt{\textbackslash footnotetext} entry in the the correct place in the list.}

Magnetic Fluids are dipolar fluids with numerous applications  \cite{FF_Apll1, FF_Apll2}. These materials are based on monodomain magnetic nanoparticles (NPs) dispersed in a liquid carrier with a stabilization against aggregation performed either with a steric coating or with an electrostatic double layer (in polar carriers) \cite{FF_Gene1, FF_Gene2}. When the dipolar interaction is dominant, the nanoparticles self-assemble under field into anisotropic structures \cite{PRE Philipse Wiedenmann}, driven by the formation of dipolar chains with NPs at contact and attracted together to produce column formation (like in non Brownian electrorheologic fluids \cite{Electrorheo_Fluid}). On the contrary when the interparticle repulsion is dominating as in the present work, the dipolar interaction (which can be tuned by applying a magnetic field through the progressive alignment of the NP permanent magnetic moment) only modulates the liquid-like structural organization of the Magnetic Fluid \cite{Guill_JPCB06, Wagner_JCP06}. Concentrated Magnetic Fluids then present an under-field structure with anisotropic  features of the structure factor $S(\vec q)$,  \cite{FloGa_PRE02, FloGa_JPhysCondMat03, Guill_JPCB06, Guill_MagnHydr12} both on macroscopic scales (at low scattering vectors $\vec q$) and on the local scale close to $S(\vec q)$ maxima at $\vec q^{~max}$. The low $q$ anisotropy of $S(\vec q)$ is well explained. It has been extensively described with a mean field model \cite{Cebers_MagnHydr82, Bacri_PRL95, FloGa_PRE02} and with a mean spherical model \cite{Morozov96}. We focus here on the anisotropy of $S(\vec q)$ around its maximum and first of all, on the anisotropy of $q^{max}$ itself. This corresponds to the lengthscale of the cages made by the first neighbours, which entrap the NPs. Numerical simulations of dipolar soft sphere fluids \cite{Ivanov_JMMM11} predict in that case an anisotropic local organization of the NPs more structured along the field than perpendicularly to the field, forming column-like structures inside the liquid carrier. Previous experimental studies \cite{FloGa_PRE02, FloGa_JPhysCondMat03, Guill_JPCB06} performed by Small Angle Neutron Scattering (SANS) have evidenced an opposite behavior with a structure more marked in the direction perpendicular to the field. However these SANS experiments were not able to evidence any anisotropy of the mean interparticle position (anisotropy of $q^{max}$). Thanks to the much better spatial resolution of Small Angle X-ray Scattering (SAXS), the measurements presented here show a clear anisotropy of $q^{max}$. We present here a simple model based on the field-induced magnetostriction of the NP cages at constant volume. It allows to reproduce the anisotropy of $q^{max}$ and $S(q^{max})$, experimentally measured by SAXS.\\

\indent After giving the details of our experiment in part~2, we present in part~3 the zero field experimental results  concerning NP cages and bulk modulus $B_{0}$ of the magnetic fluids in zero field, together with the NP mean quadratic displacement in their cage. The field-induced results are then presented in part~4 before describing our theoretical model in part~5. It is then compared with respect to the experimental results in part~6. Its limitations are then discussed in part~7.

%%%%
\section{\label{sec:} Materials and method}

\subsection{\label{sec:Samples} Samples}

The Magnetic Fluids studied here are prepared, as described in \cite{Guill_JPCB06,Wanders_PRE09}.
They consist of aqueous dispersions of maghemite ($\gamma-Fe_{2}O_{3}$) nanoparticles (typically 10 nm in diameter) coated with citrate molecules to ensure a negative surface charge at $pH=7$ ($\sim 2 e^{-}/nm^{2}$) \cite{Emma_JChemPhys99}. The interaction between NPs is composed of (i) van der Waals attraction, (ii) electrostatic repulsion, that can be screened by the presence of free ions in the solution, and (iii) anisotropic dipolar interaction between the magnetic moment $\vec \mu$ of the NPs which are magnetic monodomains. A chemical control of the dispersions allows increasing the weight of electrostatic repulsion leading to the colloidal stability of the Magnetic Fluid \cite{Guill_MagnHydr12, Cousin_PRE03}.

The NP size polydispersity at the end of the chemical synthesis is reduced thanks to a size-sorting process \cite{Cousin_PRE03}. We keep the largest nanoparticles among the synthesis batch in order to obtain a rather large magnetic dipolar interaction. High NP concentrations at fixed ionic strength are obtained thanks to an osmotic stress \cite{Guill_JPCB06,Wanders_PRE09}. The citrate species adsorbed on the NPs are in equilibrium with free citrate species. The concentration of these free species $[cit]_{free}$ is fixed in the dialysis bath in order to keep a strong enough electrostatic repulsion to maintain the colloidal stability of the Magnetic Fluid under an external magnetic field. We therefore do not observe here any demixing in two phases \cite{Cousin_PRE03, Lacoste-Lubensky, les croates} but we can observe an anisotropy of interaction in the Small Angle Scattering spectra \cite{Guill_JPCB06}. Most of the samples probed here are fluid samples; However one (sample B, see Table 1) is close to the colloidal glass transition \cite{Wanders_PRE09, Guill_JPhysCondMat06, Aymeric_EPL06}. The NPs volume fraction $\Phi$ in each sample is given in Table 1 - together with $[cit]_{free}$ and the NP magnetic characteristics obtained by magnetization measurements at room temperature. $d^{\it0}_{NP}$ and $s$ are the median NP  diameter and the polydispersity of the log-normal distribution of NP magnetic diameters $d_{NP}$ respectively. They are obtained from the adjustment of MF magnetization at low volume fraction by a Langevin function weighted by the log-normal distribution of diameters \cite{FF_Apll1}, with a NP saturation magnetization $m_{s}$ = 3.10$^{5}$ A/m. $M_{_{MF}}^{sat}=m_{s}\Phi$ is the saturation magnetization of the different MF samples at volume fraction $\Phi$. The dipolar interaction parameter $\Psi_{dd}$ (characteristic of the NPs and defined as  $\Psi_{dd}=\gamma/\Phi=\frac{\mu_{0}}{k_{B}T} m_{s}^{2}\frac{\pi}{6} d_{NP}^{3}$ - see Annex I) is experimentally determined in low fields by the measurement of the initial magnetic susceptibility at low volume fraction \cite{Guill_JPCB06, FloGa_PRE02, FloGa_JPhysCondMat03, Guill_MagnHydr12, Bacri_PRL95}.

 \begin{table}[h]
\small
 \caption{\ Chemical and magnetic characteristics of the samples (see text). They are in the Fluid phase except sample B which is a freshly prepared Glass Forming sample (see \cite{Wanders_PRE09})}
  \label{Table1} 
  \begin{tabular*}{0.48\textwidth}{@{\extracolsep{\fill}}llllllll}
\hline
Samples &  $[cit]_{free}$&$d^{\it0}_{NP}$&   $s$ & $\Psi_{dd}$ & $\Phi$  & $M_{_{MF}}^{sat}$ \\
 & M & nm & & &  & kA/m \\
\hline
A - Fluid
 & 0.03 & 9.5 & 0.35 & 56 & 17.5 \% & 53\\

B - Glassy
 & 0.03  & 9.5 & 0.35 & 56 & 30 \% & 90\\

C - Fluid
 & 0.03  & 9.8 & 0.25 &  34 & 12.5 \% & 38\\

 D - Fluid
 & 0.01  & 8.5 & 0.35 &  44 & 16 \% & 48\\

 E - Fluid
 & 0.01  & 9.5 & 0.4 &  80 & 13 \% & 39\\ 
\hline
\end{tabular*}
\end{table}
%%%
%
% UNDER FIELD PATTERN & STRUCTURE FACTORS
%\begin{figure}[h]
\begin{figure*}
\begin{center}
\includegraphics*[width=12.2 cm]{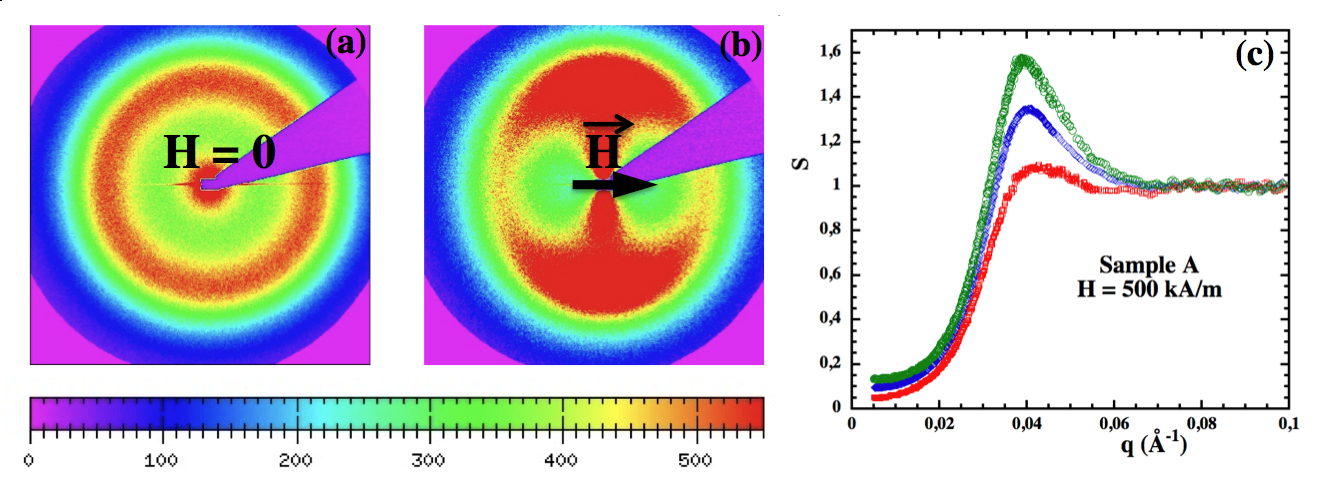} 
\caption{SAXS Patterns of sample A with $\Phi = 17.5 \%$ in zero field (Fig.~1-a) and under magnetic field (Fig.~1-b at $H$ = 500 kA/m). Their associated structure factors are presented in Fig.~1-c ; middle curve $S_{0}(q)$ from Fig.~1-a; Field induced structure factors are deduced from the analysis of Fig.~1-b pattern over $\pm 10^{\circ}$ sectors respectively centrered along the axis $\vec q \parallel \vec H$ ($S_{\parallel}(q)$ - lower curve) and the axis  $\vec q \perp \vec H$ ($S_{\perp}(q)$ - upper curve).}
 \label{Fig1}
 \end{center}
  \end{figure*}
 %
%% Table 1
%
\subsection{\label{sec:Exp} Experimental details}
SAXS experiments are realized at the ID02 beamline at the European Synchrotron Radiation Facility (Grenoble - France) using 12 keV X-rays and two sample-to-detector distances. It gives access to scattering vectors in the range $5 \cdot 10^{-3}~\mathrm{\AA^{-1}} \leqslant q \leqslant  2\cdot 10^{-1}~\mathrm{\AA^{-1}}$ with an accuraccy $\delta q = \pm~5\cdot 10^{-5}~\mathrm{\AA^{-1}}$. The intensity is detected on a FReLON CCD. The samples are prepared in 1 mm diameter quartz-capillaries. A uniform magnetic field $H$, normal to the X-ray beam in the horizontal scattering plane, can be applied with an intensity ranging from 0 to 800 kA/m. The spectra are analyzed as in \cite{Guill_JPCB06} leading in zero field to the structure factor $S_{0}(q, \Phi)$ from a radial analysis of the isotropically scattered intensity. It presents a maximum at intermediate $q$ marking the most probable interparticle distance in the isotropic dispersion. Under-field the scattering pattern is anisotropic. The intensity is thus analyzed over angular sectors of $20^\circ$ width both along the field leading to $S_{\parallel} (q, \Phi, H)$ or perpendicular to the field leading to $S_{\perp}(q, \Phi, H)$ (see \cite{Guill_JPCB06}). This is illustrated by Figure 1 which presents the SAXS patterns of sample A (see Table 1) in zero field and under a field of 500 kA/m, together with the associated structure factors. Under field, the structure factors $S_{\parallel}(q)$ and $S_{\perp}(q)$ are deduced from the analysis of the pattern over sectors respectively centrered along the axis $\vec q \parallel \vec H$ and the axis  $\vec q \perp \vec H$. In both directions the liquid-like structure factor presents a maximum respectively located at $q^{max}_{\parallel}$ and $q^{max}_{\perp}$.

In the following part, we first compare the experimental results in zero field to SANS data \cite{Guill_JPCB06,Elie_ConfJap, Th_Guill,Th_Elie} obtained in similar conditions either on PAXY at reactor Orph\'ee - LLB - Saclay - France or on D22 at ILL - Grenoble - France.

\section{\label{sec:H=0 results} Experimental results in zero field}

We first focus on the zero-field elastic properties of the magnetic fluid obtained at low $q$'s and on the structure factor shape determined over the whole $q$-range.
\subsection{Bulk modulus in zero field}
Magnetic fluids are compressible colloidal dispersions of magnetic NPs and the bulk modulus $B_{0}$ of the NP's system can be determined experimentally by SAXS and SANS. It is related to the measured isothermal compressibility $\chi_{T,0}$ of the Magnetic Fluid in zero field \cite{FloGa_PRE02, Guill_JPCB06} through

\begin{equation}
\chi_{T,0}=S_{0}(q=0)=\frac{1}{d_{0}^{3}}\frac{k_{B}T}{B_{0}}
\label{Eq:1}
\end{equation}
where $d_{0}$ is the mean interparticle distance. When the system is in the strongly repulsive regime, then $q_{0}^{max}$ scales as $\Phi^{1/3}$ and $d_{0}=2\pi/q_{0}^{max}$. Introducing the mean quadratic displacement $\sigma_{0}$ of the nanoparticles, we obtain :
\begin{equation}
\sigma _{0}^2=\frac{k_{B}T}{d_{0} B_{0}}
\label{Eq:2}
\end{equation}
with
\begin{equation}
\chi_{T,0}= \Big(\frac{\sigma _{0}}{d_{0}}\Big)^2.
\label{Eq:3}
\end{equation}
% 
% Mesures a H=0
\begin{figure}[h!]
\begin{center}
\includegraphics*[width=5.25 cm]{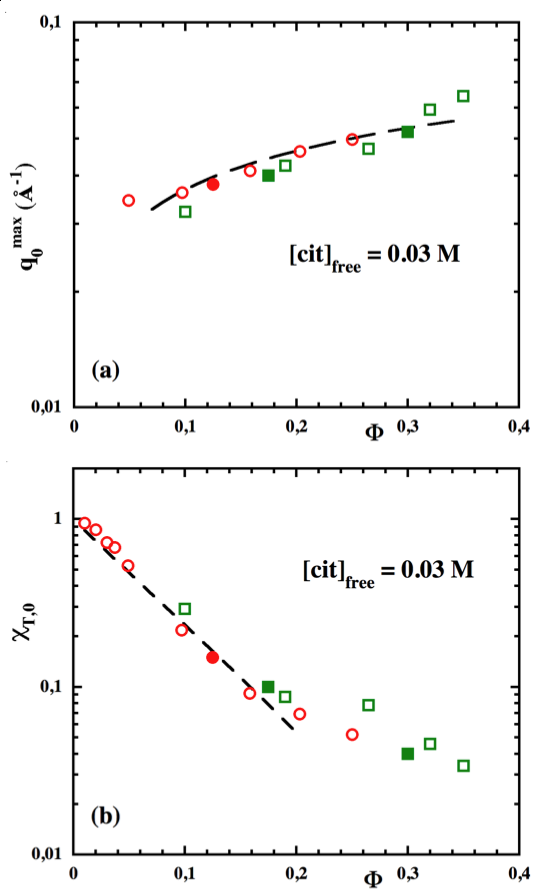} 
\caption{$\Phi$-dependence of $q_{0}^{max}$ (Fig.~2-a) and $\chi_{T,0}$ (Fig.~2-b) as deduced either from SANS (open symbols) or SAXS measurents (full symbols) for the series of samples from \cite{Guill_JPCB06,Elie_ConfJap, Th_Guill,Th_Elie}, which are based on same magnetic NPs and prepared at same $[cit]_{free}$ as samples A, B and C of Table 1; Squares correspond to same NPs as samples A $\&$ B.; Discs same NPs as sample C. Dashed line of (Fig.~2-a) : adjustment of $q_{0}^{max}(\Phi)$ by $q_{0}^{max}$ = 0.1($\Phi$/2)$^{1/3}$ corresponding to $d_{0}=2\pi/q_{0}^{max}$ and $\Phi=\pi d_{NP}^{3}/6d_{0}^{3}$ with $d_{NP}$ = 9.8 nm}; Dashed line of (Fig.~2-b) : Adjustment of $\chi_{T,0}(\Phi)$ with the Carnahan-Starling formalism of Annex II.  
\label{Fig2}
\end{center}
 \end{figure}
\indent Figure 2 summarizes the experimental determinations of $q_{0}^{max}$ and $\chi _{T,0}$ for the samples A, B and C of Table 1 and compares them to previously published SANS results \cite{Guill_JPCB06,Elie_ConfJap, Th_Guill,Th_Elie} with the same NP's and $[cit]_{free}$. Fig.~2-a shows that the NP's system is indeed here strongly repulsive as $q_{0}^{max}$ scales as $\Phi^{1/3}$. Note in Fig.~2-b that $\chi _{T,0}$ can be described up to $\Phi \sim 20 \%$ by the Carnahan-Starling formalism developped in Annex II while replacing $\Phi$ by an effective volume fraction $\Phi_{eff}$ taking in account the screening length of the electrostatically charged NP's. Above $\Phi$ of the order of 12$\%$, $\chi _{T,0}$ becomes smaller than 0.1 and thus the NPs dispersion is only weakly compressible in zero field.\\
% sigma_0 et B_0
\begin{figure}[h]
\begin{center}
\includegraphics*[width=6.2cm]{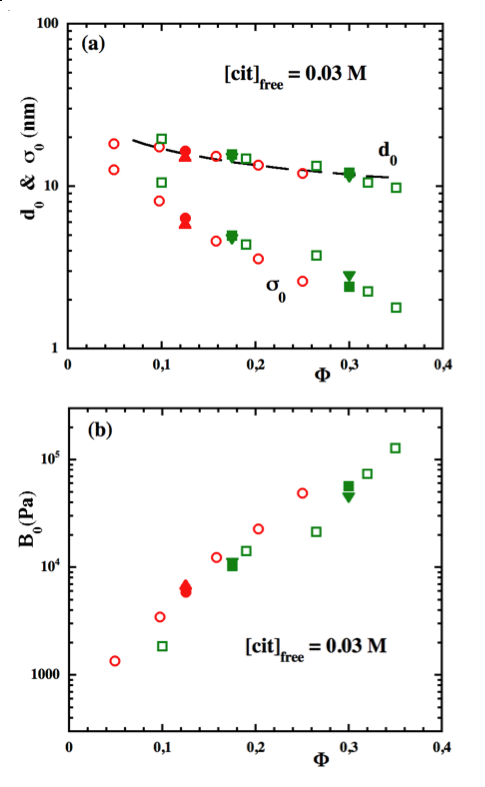} 
\caption{$\Phi$-dependence of $d_{0}$, $\sigma _{0}$ (Fig.~3-a) and $B_{0}$ (Fig.~3-b) deduced from the data of Figure 2 - same symbols as in Figure 2; $\bar d$, $\bar\sigma _{0}$ (Fig.~3-a) and $\bar B_{0}$ (Fig.~3-b) deduced from the analysis of SAXS profiles $S(q)$ by Eq.~(4) for samples A, B (full downward triangles) and C (full upward triangle). Dashed line in Fig.~3-a corresponds to $d_{0}$(nm) =  9.8.$(\pi/6\Phi)^{1/3}$ (same adjustment as in Fig.~2-b).} 
  \label{Fig3}
 \end{center}
  \end{figure}
\indent Figure 3 presents the $\Phi$-dependence of $d_{0}$, $\sigma _{0}$ and $B_{0}$ deduced from Fig.~2 using Eqs.~(1), (2) and (3). While both $d_{0}$ and $\sigma _{0}$ progressively decrease with $\Phi$, $B_{0}$ increases by two orders of magnitude over the whole range of volume fractions $\Phi$. Let us note in Fig.~3-a that if at low $\Phi$ the interparticle distance $d_{0}$ and the mean quadratic displacement $\sigma_{0}$ of NPs are of the same order, it is no more the case at large $\Phi$'s. When the system is becoming glassy (here for $\Phi \gtrsim 30\%$ \cite{Guill_JPCB06, Wanders_PRE09}) we observe that $\sigma_{0}/ d_{0}\lesssim 0.2$ but also that $B_{0}$ does not present any strong discontinuity. It smoothly becomes of the order of $10^{5}$ Pa, which is also the order of magnitude of the osmotic pressure of the NP's system. \\

The values of $d_{0}$, $\sigma _{0}$ and $B_{0}$ resulting from the analysis of the SAXS spectra of Table 1 samples are summarized in Table 2. Samples with similar volume fractions (A $\&$ D on the one hand and C $\&$ E on the other hand - see Table 1) present close $B_{0}$ values, independently of $[cit]_{free}$.\\
The $B_{0}$ value of the glass-forming sample B can be compared to that obtained for glass-forming systems based on different microscopic objects. Indeed we can rewrite Eq.~(1) as $\chi_{T,0}=\frac{1}{(d_{NP}^{\it0})^{3}}\frac{\Phi k_{B}T}{B_{0}}$. At equivalent $ \frac{1}{\Phi}\frac{\chi_{_{T,0}}}{ k_{B}T}$, $B_{0}$ should scale as the inverse of the volume of the dispersed objects. The value $B_{0}$ = 6.2 10$^{4}$ Pa found here scales well with the elastic modulus of the suspension of micron size silica particle close to the glass transition $0.1~Pa$ given in \cite{Weitz_Science07} since the ratio of the characteristic sizes for these two systems is approximately equal to 100.

%% Table 2
%\begin{center}
\begin{table*}
\small
 \caption{\ Characteristics of the samples deduced from SAXS measurements; $d_{0}$ is the mean interparticle distance in zero field and $\chi_{T,0}$ the experimental isothermal compressibility; $\sigma_{0}$ and $B_{0}$ are experimentally deduced from $d_{0}$ and $\chi_{T,0}$ using Eqs.~(1), (2) and (3); $\bar d$, $\bar \sigma_{0}$ and $\bar B_{0}$ are deduced from the adjustment of S(q) with Eq.~(4); The under-field interparticle distances $\tilde d_{\parallel}=d_{\parallel}(H_{max})$ and $\tilde d_{\perp}=d_{\perp}(H_{max})$ are deduced from $q_{max}$ values at maximum field; $K_{H}^{el}$ is deduced from the under-field model of part~5 (Eq.~(20), Fig.~6) and the magnetic characteristics of table 1}
  \label{Table2} 
  \begin{tabular*}{\textwidth}{@{\extracolsep{\fill}}llllllllllll}
\hline
Sample  &  $d_{0}$ & $\chi_{T,0}$ & $\sigma _{0}$ & $B_{0}$ &$\bar d$  &$\bar \sigma _{0}$ & $\bar B_{0}$ & $\tilde d_{\parallel}$ & $\tilde d_{\perp}$ & $K_{H}^{el}$ \\
 & (nm)&  & (nm) & (Pa) & (nm) &  (nm)& (Pa) &  (nm) & (nm)  & (Pa)  \\
\hline
A
& 15.7 & 0.1 & 5.0 & 1.0 10$^{4}$& 14.9 &4.7 & 1.1 10$^{4}$& 14.8 & 16.2 & 2.6 10$^{4}$\\

B
 & 11.7&0.04&2.3&6.2 10$^{4}$&11.3 &2.8 & 4.4 10$^{4}$& 11.1 & 12 &9.4 10$^{4}$\\

C
& 16.7 &0.15 & 6.5 & 5.7 10$^{3}$& 15.2&5.9 & 6.9 10$^{3}$& 16.2 & 17 & 2.7 10$^{4}$\\

D
& 17.6&0.06&4.3&1.2 10$^{4}$&17&4.8 &1.0 10$^{4}$& 16.1 & 18.3 & 1.7 10$^{4}$\\ 

E
& 22.2 &0.05 &5.0&7.3 10$^{3}$&21,4 &6.0 & 5.0 10$^{3}$& 20.9 & 22.8 & 1.4 10$^{4}$ \\ 
\hline
\end{tabular*}
\end{table*}
%\end{center}
%%
%
% ZERO FIELD STRUCTURE FACTOR + FIT
\begin{figure}[h]
\begin{center}
\includegraphics*[width=6cm]{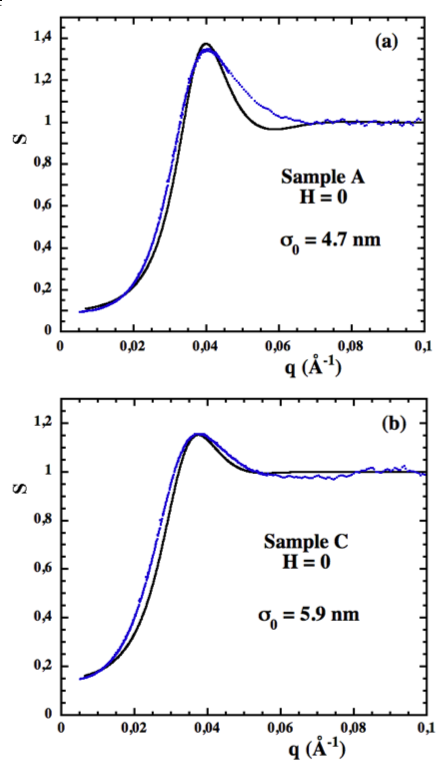} 
\caption{Zero-field S(q) profiles of sample A (Fig.~4-a) and sample C (Fig.~4-b) adjusted by Eq.~(4) (solid line) with respectively $(\bar \sigma_{0}/d_{0})^2$ = 0.09 and 0.125.} 
  \label{Fig4}
 \end{center}
  \end{figure}
%%%%
\subsection{Paracrystal structure in zero field}
\indent Another way to determine the NP's mean quadratic displacement and the bulk elastic modulus is to adjust the zero-field structure factor $S_{0}(q)$ of the dispersions by the following expression standardly used for colloidal dispersions of (monodisperse) nanoparticles \cite {Matsuoka_PRB87} :\begin{equation}
S(q)=\frac{\sinh{(\bar \sigma_{0}^{2}q^{2}/2)}}{\cosh{(\bar \sigma_{0}^{2}q^{2}/2)}-\cos{(q\bar d)}}
\label{Eq:4}
\end{equation}
where $\bar d$ is a parameter, which equals $d_{0}=2\pi/q_{0}^{max}$ only if $\bar \sigma_{0}/\bar d\ll 1$.  We call $\bar \sigma_{0}$ and  $\bar B_{0}$ the NP's mean quadratic displacement and the bulk elastic modulus determined by this method. Eq.~(4) has been used for example for describing the structure of latex solutions \cite{Matsuoka_PRB87} or the fluctuations of multivalent ions adsorbed on a linear polyelectrolyte chain \cite{Cebers_PRL06, livre-Lubensky}. \\
\indent In the present case, the condition $\bar \sigma_{0}/\bar d\ll 1$ is not fulfilled and $\bar d$ has to be fitted. Therefore $S(q)$ is rewritten as a fonction  of the two parameters $\bar \sigma_{0}/\bar d$ and $q\bar d$ which are both adjusted, $\bar \sigma_{0}/\bar d$ controlling the shape of $S(q)$ and $q\bar d$ controlling the position of the $S(q)$ maximum. The values of $\bar \sigma_{0}$ and $\bar B_{0}$ obtained with such fits are summarized in Table 2 and plotted in Fig.~3. They are rather close to the values previously obtained from $\chi_{T,0}$ and $q_{0}^{max}$. Fig.~4 illustrates the quality of the adjustment of $S_{0}(q)$ by Eq.~(4) for samples A and C.\\
 Despite the fact that Eq.~(4) does not take into account the polydispersity of the NPs, we can show a very good self-consistency between the determinations of various parameters for system in zero field from either the $S(q)$ profile adjustment or from the compressibility determinations at low $q$'s. \\

\section{\label{sec: under-H results} Under-fields results}

The scattering profiles present a strong anisotropy when applying an external magnetic field (cf Fig.~1). We focus here on the analysis of the structure factor in the directions parallel and perpendicular to the field.\\

First of all we find that experimentally $q^{max}_{\parallel}(H)$ is always larger than $q^{max}_{\perp}(H)$, meaning that the interparticle distance $d_{\parallel}=2\pi/q^{max}_{\parallel}$ is always smaller than $d_{\perp}=2\pi/q^{max}_{\perp}$. If the cage formed by the first neighbours around a given nanoparticle is approximated by an ellipsoid, this means that the zero-field spherical cage always deforms under-field as an oblate ellipsoid. Moreover at the first order, the deformation of this cage occurs at constant volume. Indeed experimentally the ratio $d_{\parallel}(H)d_{\perp}^{2}(H)/d_{0}^{3}$ is found equal to 1 $\pm 0.02$ for every applied field and whatever the sample.\\

Figure 5 shows for sample A the field dependence of $q^{max}_{\parallel}$ and $q^{max}_{\perp}$ and that of $S^{max}_{\parallel}$ and $S^{max}_{\perp}$. Their under-field anisotropy goes in opposite ways and saturates in high fields. The values of $\tilde d_{\parallel}=d_{\parallel}(H_{max})=2\pi /q_{\parallel}^{max}(H_{max})$ and $\tilde d_{\perp}=d_{\perp}(H_{max})=2\pi /q_{\perp}^{max}(H_{max})$ are reported in Table 2.

%
% FIELD DEPENDENCE OF Qmax para & perp
\begin{figure}[h]
\begin{center}
\includegraphics*[width=6cm]{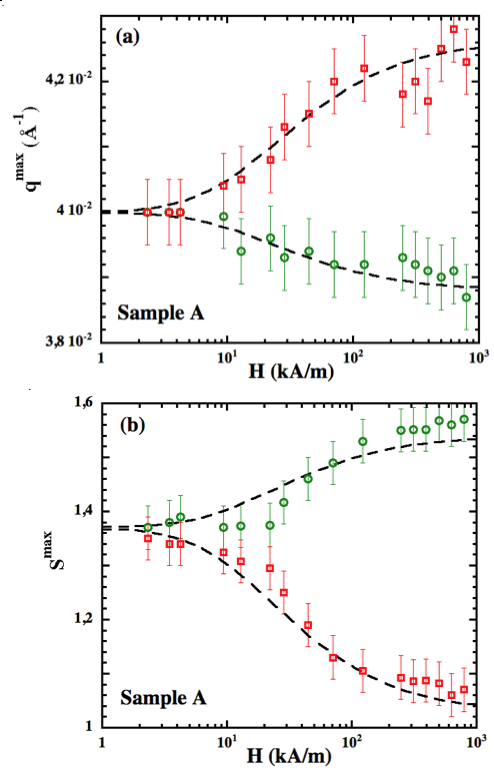} 
\caption{Field dependence of $q^{max}_{\parallel}$ and $q^{max}_{\perp}$ (Fig.~5-a) and of $S^{max}_{\parallel}$ and $S^{max}_{\perp}$ (Fig.~5-b) for (fluid) sample A (same sample as in Fig.~1 - see characteristics in Table 1). Symbols :  squares in direction parallel to the applied field and circles in the perpendicular direction. The dashed lines (see part~6) correspond to best adjustments of the data to Eqs. (20) and (22) using for $M_{_{MF}}$ the effective mean field model of Annex I with $d_{NP}=11.5$ nm. It leads to $K^{el}_{H} = 2.6$ $10^{4}$ Pa.} 
\label{Fig5}
\end{center}
\end{figure}

\section{\label{sec:TheoModel} Theoretical model}

In these dispersions, each NP bears a permanent magnetic dipole $\vec\mu$. If the dispersion is dilute, the magnetization $M_{_{MF}}$ can be described by a Langevin formalism and results from the progressive reduction of the orientational fluctuations of the magnetic dipole $\vec\mu$ around the direction of the applied field $\vec H$. In a concentrated dispersion, these dipoles interact together through magnetic dipolar interaction. Between two parallel dipoles $\vec \mu$ at distance $d$ from each other, this dipolar interaction is anisotropic and manifests itself as attractive along the direction of the magnetic dipoles and repulsive in the perpendicular direction. It can be written at the first order as :
\begin{equation}
u^{\parallel}_{dd}=-\frac{\mu_{0}\mu^{2}}{2\pi d^{3}}\quad\text{and} \quad u^{\perp}_{dd}=\frac{\mu_{0}\mu^{2}}{4\pi d^{3}}
\label{Eq:5}
\end{equation}
The model of Annex I (from \cite{Bacri_PRL95,FloGa_PRE02}) describes the $H$-dependence of $M_{_{MF}}$ in concentrated magnetic fluids. Under an applied field, the dipolar interaction induces a uniaxial stress between NP's, leading to a magnetostriction at constant volume, without compression. Macroscopically the magnetostrictive contribution to the energy density of the magnetic colloid may be obtained considering the total field acting on the dipoles, as in the mean field model of  \cite{Cebers_MagnHydr82}. This latter model describes well the effect of the dipolar interaction on the thermodynamic properties of the magnetic colloids that are measured in the limit of $q \rightarrow 0$ \cite{Guill_JPCB06, Bacri_PRL95,FloGa_PRE02}. On a more local scale, each NP in a concentrated Magnetic Fluid can be seen as entrapped in a cage constituted by its first neighbouring NPs. Because of the field-induced magnetostriction, the cage around a magnetic NP, while keeping a constant volume, becomes anisotropic with a dimension $d_{\parallel}$ smaller than $d_{\perp}$. Considering the experimental observations described in part~4, we approximate the cage by an oblate ellipsoid of excentricity $e$ and constant volume $V_{cage}=\frac{\pi}{6}d_{\parallel}d_{\perp}^{2}\simeq \frac{\pi}{6}d_{0}^{3}$, with its symmetry axis along its magnetization $\vec M_{cage}$. Inside the ellipsoidal cage, $\vec M_{cage}$ is assumed to be homogeneous and each cage is associated to a magnetic moment $\vec m$, mean projection along $\vec H$ of the fluctuating moment $\vec \mu$ in the cage. $\vec m$ is field dependent with $m(H)=M_{_{MF}}(H)d_{\parallel}d_{\perp}^{2}= M_{cage}(H)V_{cage}$. The energy per particle associated to the demagnetizing field is then :
\begin{equation}
E=-\frac{\mu_{0}}{2}N(e)M_{cage}^{2}V_{cage}
\label{Eq:6}
\end{equation}
where $\mu_{0}$ is the vacuum permeability and $N(e)$ the demagnetizing factor of the ellipsoidal cage. At small deformations ($z$ axis is along the direction of the magnetization, $x,y$ axes are perpendicular to it), the eccentricity of the oblate ellipsoid is :
\begin{equation}
e^{2}=(u_{x,x}+u_{y,y})-2u_{z,z}
\label{Eq:7}
\end{equation}
where $\vec u=(u_{x},u_{y},u_{z})$ is the displacement vector with respect to $H=0$ position, the second subscript of $u_{i,i}$ denoting the partial derivative with respect to the corresponding variable of $u_{i}$. Accounting for the expression of the demagnetizing factor of an ellipsoid of small eccentricity :
\begin{equation}
N(e)=\frac{1}{3}(1+\frac{2}{5}e^{2}),
\label{Eq:8}
\end{equation}
and neglecting the constant term, we obtain the magnetostriction energy :
\begin{equation}
E_{m}=\frac{4 \mu_{0}}{5 \pi}M_{_{MF}}^{2}d_{0}^{3}(u_{z,z}-\frac{1}{2}(u_{x,x}+u_{y,y})).
\label{Eq:9}
\end{equation}
In the absence of overall under-field compression ($\Delta V_{cage}/V_{cage} =0$ and thus $\Sigma_{i}u_{i,i}=0$ whatever $H$) $u_{z,z}=-(u_{x,x}+u_{y,y})$, Eq.~(9) has thus a form equivalent to that of the magnetostriction energy considered in \cite{Landau_Lifshitz} in the case of the ferromagnetics. For the present uniaxial deformation along $z$ axis let us note $u_{z,z}=(z_{j+1}-z_{j}-d_{0})/d_{0}$, with $z_{j}$ the average position of the $j^{th}$ particle along $z$ axis. We obtain for the energy per particle 
\begin{equation}
E_{m}=\frac{6\mu_{0}}{5\pi} M_{_{MF}}^{2}d_{0}^{2}(z_{j+1}-z_{j}-d_{0}).
\label{Eq:10}
\end{equation}
Eq.10 is close to the energy $U^{\parallel}_{dd}$ of two parallel dipoles with the radius vector along their magnetic moment $\vec m$ :
\begin{equation}
U^{\parallel}_{dd}=-\frac{\mu_{0}m^{2}(H)}{2\pi (z_{j+1}-z_{j})^{3}}\simeq-\frac{\mu_{0}m^{2}}{2\pi d_{0}^{3}}+\frac{3\mu_{0}m^{2}}{2\pi d_{0}^{4}}(z_{j+1}-z_{j}-d_{0})
\label{Eq:11}
\end{equation}
Here the magnetic moment $\vec m$ being field-dependent with $m(H)=d_{0}^{3}M_{_{MF}}(H)$, we see that the magnetostrictive part of Eq.~(11), except for a coefficient 0.8, coincides with the general Eq.~(10). Further on we use the estimate given by Eq.~(11).

In concentrated fluid samples the under-field interaction at mean distance $d$ between the colloidal particles is related to the elastic deformation of their surroundings. Denoting the mean quadratic displacement of the particles around their mean position under-field by $\sigma^{2}_{H}$ and assuming that it does not depend on the field value nor on its direction, we have for the global energy of a sequence of $N$ particles along the field direction :
\begin{eqnarray*}
\frac{E_{H}}{k_{B}T}=\sum^{N-1}_{j=0}\Bigl( \frac{1}{2\sigma_{H}^{2}} (z_{j+1}-z_{j}-d_{0})^{2}\quad \quad\quad\quad\quad\quad \text{}
\end{eqnarray*}
\begin{equation}
\quad \quad\quad\quad\quad\quad\quad\quad+\frac{3\mu_{0}m^{2}(H)}{2\pi k_{B}Td_{0}^{4}}(z_{j+1}-z_{j}-d_{0})\Bigr).
\label{Eq:12}
\end{equation}
Providing that $3\mu_{0}m^{2}(H)\sigma_{H}^{2}\ll 4\pi k_{B}Td_{0}^{5}$, the effective energy of the particle interaction in the sequence of $N$ particles can be rewritten in these terms :
\begin{equation}
\frac{E_{H}}{k_{B}T}=\frac{1}{2\sigma_{H}^{2}}\sum^{N-1}_{j=0}(z_{j+1}-z_{j}-d_{\parallel})^{2}+const
\label{Eq:13}
\end{equation}
with
\begin{equation}
d_{\parallel}=d_{0}\Bigl(1-\frac{3\mu_{0}m^{2}(H)\sigma_{H}^{2}}{2\pi k_{B}Td_{0}^{5}}\Bigr)
\label{Eq:14}
\end{equation}
giving the renormalization of the mean distance between the particles due to the magnetostriction. Eq.~(13) leads to the under-field elastic energy per unit volume 
\begin{equation}
e_{elas,H}=\frac{1}{2}\frac{k_{B}T}{\sigma_{H}^{2}d_{0}}u^{2}_{z,z}
\label{Eq:15}
\end{equation}
and gives, in a way similar to Eq.~(2), the following estimate for the under-field elastic modulus $K^{el}_{H}$ : 
\begin{equation}
K^{el}_{H}=\frac{k_{B}T}{\sigma_{H}^{2}d_{0}}.
\label{Eq:16}
\end{equation}
$K^{el}_{H}$ is analogous to a Young modulus in this anisotropic elastic medium. It is different from the compression modulus $B_{0}$ measured at $H$ = 0. We thus obtain :
\begin{equation}
\frac{d_{\parallel}}{d_{0}}=1-\frac{3\mu_{0}M_{_{MF}}^{2}}{2\pi K^{el}_{H}}.
\label{Eq:17}
\end{equation}
We note that the relative decrease of the mean distance between the particles in the direction of the magnetization is field-dependent through the field-dependence of $M_{_{MF}}$ and that it does not depend explicitly on $d_{0}$. It might depend on $d_{0}$ through the $d_{0}$-dependences of the elastic modulus $K^{el}_{H}$ and magnetization $M_{_{MF}}$. 

Let us note that the condition to write Eqs.~(14-15) now reads $3\mu_{0} M_{_{MF}}^{2}<<4 \pi K^{el}_{H}$.

In a similar way considering the interaction energy of two parallel dipoles with the radius vector perpendicular to their direction
\begin{equation}
U^{\perp}_{dd}=\frac{\mu_{0}m^{2}}{4\pi (x_{j+1}-x_{j})^{3}}\simeq\frac{\mu_{0} m^{2}}{4\pi d_{0}^{3}}-\frac{3\mu_{0} m^{2}}{4\pi d_{0}^{4}}(x_{j+1}-x_{j}-d_{0}),
\label{Eq:18}
\end{equation}
we obtain the mean distance between particles in the direction perpendicular to the magnetization of the sample
\begin{equation}
\frac{d_{\perp}}{d_{0}}=1+\frac{3\mu_{0}M_{_{MF}}^{2}}{4\pi K^{el}_{H}}.
\label{Eq:19}
\end{equation}

%%%%
\section{\label{subsec:CompExp} Test of the model - Comparison to experiment}

At the highest fields of the experiment $M_{_{MF}}$ is saturated whatever the sample. $K^{el}_{H}$ can thus be easily evaluated from the relation $K^{el}_{H}=9\mu_{0}{M_{_{MF}}^{sat}}^{2}d_{0}/4\pi(\tilde d_{\perp}-\tilde d_{\parallel})$ deduced from Eqs.~(17) and (19).
The values of $K^{el}_{H}$ (summarized in Table 2) are obtained for each sample from the adjustment of the global $H$-dependences of $q^{max}_{\parallel}$ and $q^{max}_{\perp}$ experimentally measured with the following expression : 

\begin{eqnarray*}
q^{max}_{\parallel}(H)=q^{max}_{0}/\Bigl(1-\frac{3\mu_{0} M_{_{MF}}^{2}(H)}{2\pi K^{el}_{H}}\Bigr) \quad\text{and}\quad\quad\text{}
\end{eqnarray*}
\begin{equation}
\quad  \quad\quad \quad\quad q^{max}_{\perp}(H)=q^{max}_{0}/\Bigl(1+\frac{3\mu_{0} M_{_{MF}}^{2}(H)}{4\pi K^{el}_{H}}\Bigr)
\label{Eq:20}
\end{equation}
Eqs.~(20) are also deduced from Eqs.~(17) and (19). The $H$-dependence of $M_{_{MF}}$ in the concentrated colloids investigated here, can be calculated with the characteristics of Table 1 and by using the effective field model \cite{FloGa_PRE02, Bacri_PRL95} detailed in Annex I. 
As in the mean-field model the NP polydispersity is not taken into account, an averaged magnetic NP diameter $\sqrt[3]{<d_{NP}^{3}>}$ is used here (= 11.5 nm for sample A for example in Fig.~5). The model describes well our data; In Fig.~5-a the same value $K^{el}_{H}$ = 2.6 10$^{4}$ Pa is used for the adjustments of $q^{max}_{\parallel}(H)$ and $q^{max}_{\perp}(H)$ in both directions parallel and perpendicular to the applied field. Table 2 collects the $K^{el}_{H}$ values determined in this way for all the samples tested here. These under-field $K^{el}_{H}$ values are of the same order of magnitude as the Bulk modulus $B_{0}$ determined in zero field in part~3. They are however systematically larger by a factor of the order of a few units.\\

The under-field anisotropy of magnetic dipolar interaction between adjacent NPs in the Magnetic Fluid can be expressed in terms of SAXS determined quantities as:
\begin{equation}
\frac{U^{\perp}_{dd}-U^{\parallel}_{dd}}{k_{B}T}=\frac{3\mu_{0}}{4 \pi}\frac{M_{_{MF}}^{2}(H)d_{0}^{2}}{ k_{B}T}=\frac{K^{el}_{H}}{3k_{B}T}d_{0}^{2} (d_{\perp}-d_{\parallel})
\label{Eq:21} 
\end{equation}
Figure 6 plots $S^{max}_{\perp}(H)-S^{max}_{\parallel}(H)$ as a function of $K^{el}_{H}d_{0}^{2} (d_{\perp}-d_{\parallel})/3k_{B}T$ for the Fluid samples of Table 1. It shows that the anisotropy of $S^{max}(H)$ is proportional to the anisotropy of $U_{dd}(H)$ with $S^{max}_{\perp}(H)-S^{max}_{\parallel}(H)=\alpha(U^{\perp}_{dd}-U^{\parallel}_{dd})$ and the same experimental coefficient $\alpha\sim  0.7$ whatever $[cit]_{free}$ in the dispersion. This coefficient decreases to 0.3 in the case of the Glass Forming sample B (data not shown). Another sample comparable to sample C has been studied under magnetic field by SANS in \cite{Guill_JPCB06}. It scales the same way with $\alpha\sim  0.7$.\\

 %Analyse des resultats sous champs1
\begin{figure}[h]
\begin{center}
\includegraphics*[width=6 cm]{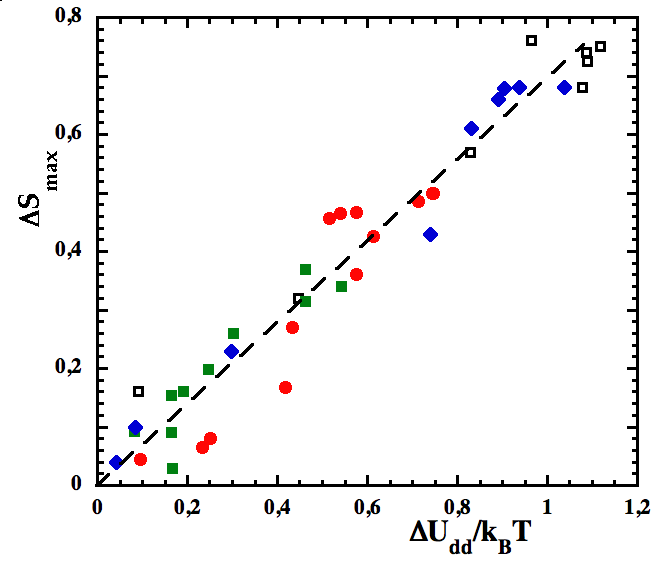} 
\caption{$S^{max}_{\parallel}-S^{max}_{\perp}$ as a function of the reduced quantity $\frac{U^{\perp}_{dd}-U^{\parallel}_{dd}}{k_{B}T}= \frac{K^{el}_{H}}{3k_{B}T}d_{0}^{2} (d_{\perp}-d_{\parallel}) $ calculated with the experimental values of $d_{0}$, $d_{\perp}$ and $d_{\parallel}$ and with $K^{el}_{H}$ from Table 2. Symbols : Sample A (closed discs), sample C (closed squares), sample D (closed diamonds), sample E (open squares). Dashed line corresponds to $S^{max}_{\parallel}-S^{max}_{\perp}$ = 0.7 $\frac{U^{\perp}_{dd}-U^{\parallel}_{dd}}{k_{B}T}$.} 
  \label{Fig6}
 \end{center}
  \end{figure}

In figure 5-b, $S^{max}_{\parallel}$ and $S^{max}_{\perp}$ are fitted with :

\begin{eqnarray*}
S^{max}_{\parallel}(H)=S^{max}_{0}-\alpha\frac{\mu_{0}M_{_{MF}}^{2}(H)}{2\pi k_{B}T} d_{0}^{3} \quad\quad\text{and}
\end{eqnarray*}
\begin{equation}
 \quad S^{max}_{\perp}(H)=S^{max}_{0} +\alpha \frac{\mu_{0}M_{_{MF}}^{2}(H)}{4\pi k_{B}T}d_{0}^{3}
\label{Eq:22} 
\end{equation}
with $\alpha$ = 0.7 and $M_{_{MF}}(H)$ adjusted as in Fig 5-a.\\

Another graphical representation of these results, strictly equivalent to the previous analysis,  consists in plotting the different ways of deducing the quantity $\frac{3\mu_{0} M_{_{MF}}^{2}}{2 \pi K^{el}_{H}}$ as Eqs. (17), (19) and (22) can be rewritten as :\\
\begin{eqnarray*}
\frac{3\mu_{0} M_{_{MF}}^{2}}{2 \pi K^{el}_{H}}=1-\frac{d_{\parallel}}{d_{0}}=2\Bigl(\frac{d_{\perp}}{d_{0}}-1\Bigr) \quad\quad\quad\quad\quad\quad\text{}
\end{eqnarray*}
\begin{equation}
 \quad\quad\quad= \frac{2k_{B}T(S^{max}_{\perp}(H)-S^{max}_{\parallel}(H))}{\alpha K^{el}_{H} d_{0}^{3}}
\label{Eq:23}
\end{equation}
%Analyse des resultats sous champs2
\begin{figure}[h]
\begin{center}
\includegraphics*[width=6cm]{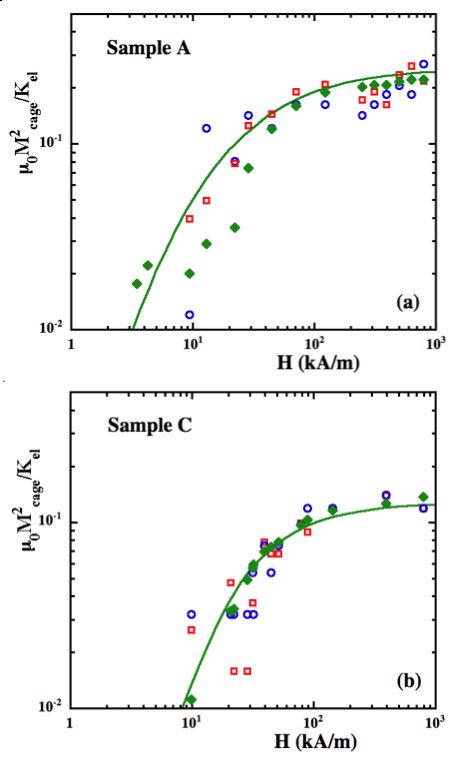} 
\caption {Field dependence of $3\mu_{0}M_{_{MF}}^{2}$/$2\pi K^{el}_{H}$ for sample A (Fig.~7-a) and for sample C (Fig.~7-b) obtained in different ways (see Eq.~(23)): (i) as equal to $1-d_{\parallel}/d_{0}$ using $q^{max}_{\parallel}$ and $q^{max}_{0}$ measurements (open squares), (ii) as equal to $2(d_{\perp}/d_{0}-1)$ using $q^{max}_{\perp}$ and $q^{max}_{0}$ measurements (open discs) and (iii) as equal to $(S^{max}_{\perp}(H)-S^{max}_{\parallel}(H))\frac{2k_{B}T}{\alpha K^{el}_{H}d_{0}^{3}}$ using $S^{max}_{\perp}$ and $S^{max}_{\parallel}$ measurements (close diamonds with $\alpha$=0.7 and $K^{el}_{H}$ from Table 2). The full lines correspond to the calculation of $3\mu_{0}M_{_{MF}}^{2}$/2$\pi K^{el}_{H}$ with $M_{_{MF}}^{2}$ calculated with the effective field model of Annex I (with $d_{NP}$ = 11.5 nm for both samples A and C) and the values of $K^{el}_{H}$ from Table 2.}
 \label{Fig7}
 \end{center}
 \end{figure}
Figure 7 compares, for sample A in Fig 7-a and sample C in Fig 7-b, the $H$-dependences of the three experimental quantities :\\
(i)  $(1-\frac{d_{\parallel}}{d_{0}})$ deduced from the $H$-dependence of $q_{\parallel}(H)$,\\
(ii)  $2(\frac{d_{\perp}}{d_{0}}-1)$ deduced from the $H$-dependence of $q_{\perp}(H)$,\\
(iii) $2k_{B}T(S^{max}_{\perp}(H)-S^{max}_{\parallel})/0.7 K^{el}_{H}d_{0}^{3}$.\\
All three superimpose with $3\mu_{0}M_{_{MF}}^{2}$/$2\pi K^{el}_{H}$ as deduced from the effective field model of Annex I with the $K^{el}_{H}$ values of Table 2 (full line in Fig 7). In this figure it is easy to verify that the condition $3\mu_{0}M_{_{MF}}^{2}<< 2\pi K^{el}_{H}$, which allows writing Eq.(14-17) is here fulfilled for samples A and B. This is true for all the samples of Table 2. This representation also clearly shows the great coherence of the experimental data in parallel and perpendicular directions both for the $q^{max}$ position of $S(q)$ peak and for the value $S^{max}$ of its maximum.\\

%%%%
\section{\label{sec:Disc} Discussion - Limits of the model}

Let us go back to the under-field profiles of $S_{\perp}(q)$ and $S_{\parallel}(q)$ of Fig.~1b from sample A at $H$ = 500 kA/m. Besides the anisotropy of $q^{max}$ and $S^{max}$, the bump of $S(q)$  obviously presents also a width anisotropy. This could eventually originate from an experimental under-field anisotropy of $\sigma$ contrarily to what is assumed in the model of parts~5 and 6.
%
% UNDER FIELD STRUCTURE FACTOR + FIT
\begin{figure}[h]
\begin{center}
\includegraphics*[width=5cm]{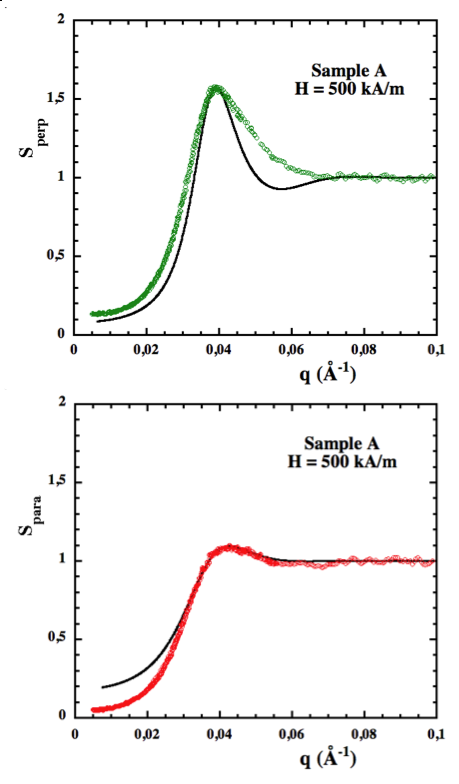} 
\caption{Under-field S(q) profiles of sample A : $S_{\perp}(q)$ in (a) and $S_{\parallel}(q)$ in (b) adjusted by Eq. 4 with respectively $(\sigma_{\perp}/d_{\perp})^2$ = 0.07 and $(\sigma_{\parallel}/d_{\parallel})^2$ = 0.14.} 
  \label{Fig8}
 \end{center}
  \end{figure}

We thus make the hypothesis that we can analyze the under-field profiles of Fig.~1b with Eq.~(4) and define $\sigma _{\parallel}(H)$ and $\sigma _{\perp}(H)$ in the two directions. Adjusting the coordinates of the maxima of $S_{\parallel}(H)$ and $S_{\perp}(H)$ while fitting them with Eq.(4), we find for sample A at $H$ = 500 kA/m (see figure 8) : $\sigma _{\parallel} = 5.65$ nm and $\sigma _{\perp} = 4.35$ nm.

%
% ANISOTROPY OF SIGMA
\begin{figure}[h!]
\begin{center}
\includegraphics*[width=6cm]{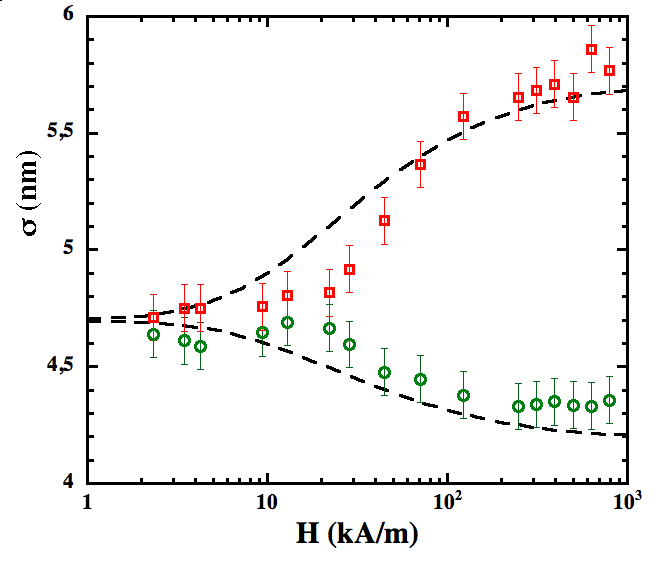} 
\caption{Under-field $\sigma$-anisotropy of sample A deduced from fitting the profiles $S_{\perp}(H)$ and $S_{\parallel}(H)$ by Eq.~(4) with an adjustement of $S^{max}_{\perp}(H)$ and $S^{max}_{\parallel}(H)$. Symbols : $\sigma _{\perp}$ (open discs) and $\sigma _{\parallel}$ (open squares) ; The dashed line corresponds to $\sigma _{\parallel}$-$\sigma _{0}$ =  $2(\sigma _{0}$-$\sigma _{\perp}) \propto M^{2}_{_{MF}}$ adjusted with a constant coefficient and with the field dependence of $M_{_{MF}}$ deduced from the effective field model of Annex I as in Figs 5 and 7a.}
  \label{Fig9}
 \end{center}
  \end{figure}

For every field $H$ we find a reasonable agreement for $3(\sigma _{0}/d_{0})^2 \sim (\sigma _{\parallel}/d_{\parallel})^2 $ + 2 $(\sigma _{\perp}/d_{\perp})^2$. The  field-dependence of $\sigma _{\parallel}$ and $\sigma _{\perp}$ is presented in Figure 9 where it is tentatively adjusted to $M^2_{_{MF}}$ (see the figure caption).

Note that the $\sigma$ values ($\sigma _{0}$, $\sigma _{\parallel}$ and $\sigma _{\perp}$) are  all of the same order of magnitude, but systematically larger than the width $\sigma_{H}=\sqrt{k_{B}T/K_{H}^{el}d_{0}}$ = 3.1 nm that is deduced from the under-field model of part~6 for this sample (using Eq.~(16) and the sample characteristics of table 2, particularly the value of $K^{el}_{H}$).

This subsidiary $H$-dependence of $\sigma$ could eventually be due to the long range dipolar interaction which introduces a supplementary force in the direction parallel to the field  \cite{Cebers_MagnHydr82, Bacri_PRL95, FloGa_PRE02, FloGa_JPhysCondMat03, Guill_JPCB06}. Indeed at very low $q$'s, thus on macroscopic scale, this supplementary force well explains the experimental structure factor anisotropy $S^{-1}_{\parallel}(q=0)$ - $S^{-1}_{\perp}(q=0)$ and its $H$-dependence. Here $S^{-1}_{\parallel}(q=0)$ - $S^{-1}_{\perp}(q=0)$ is maximum in large fields where it equals $\sim$ 11. This value is comparable to $\gamma=\Psi_{dd}\Phi$ = 9.8 (see Table 1 and Annex I) expected from refs \cite{Cebers_MagnHydr82, Bacri_PRL95, FloGa_PRE02, FloGa_JPhysCondMat03, Guill_JPCB06}. Note that the fits of $S_{\parallel}(q)$ and $S_{\perp}(q)$ by Eq.~(4) are adjusted here to the maximum of $S(q)$. As shown in Fig.~8 they are unable to model the low-$q$'s anisotropy observed in the experiments.\\
\indent We now compare sample A and sample B. They are based on the same nanoparticles but sample A is a Fluid sample \cite{Guill_JPCB06} while sample B is a (freshly prepared) Glass-forming one as in \cite{Wanders_PRE09, Wanders_BJP09} with a much larger compression modulus $B_{0}$. Table 3 compares, for these two samples, the large field anisotropies of $d$, $S^{max}$ and $\sigma$.\\

%
 %% Table III
%\begin{center}
\begin{table}[h]
\small
  \caption{\ Maximum anisotropies of $d$, $S^{max}$ and $\sigma$ determined in large magnetic fields from the scattering patterns for Samples A and B and deduced from the adjustement of the $S(q)$ profiles by Eq.~(4)}
  \label{tbl:example}
  \begin{tabular*}{0.5\textwidth}{@{\extracolsep{\fill}}lllll}
    \hline
 & & & \\
 Sample &  \large{$\frac{\tilde d_{\perp}-\tilde d_{\parallel}}{d_{0}}$} & \large{$\frac{\tilde S^{max}_{\perp}-\tilde S^{max}_{\parallel}}{S^{max}_{0}}$} & \large{$\frac{\tilde \sigma_{\perp}-\tilde \sigma_{\parallel}}{\sigma_{0}}$}\\
 & & & \\
\hline
A - Fluid
 & 8.9 \% & 37 \% & 30 \%\\

 B - Glass Forming
  & 7.7 \% & 18 \% & 3 \%\\ 
\hline
\end{tabular*}
\end{table}
%\end{center}
%%
If a comparable anisotropy of $d$ (and $q^{max}$) is observed with both samples, the anisotropy of $S^{max}$ is reduced by a factor of 2 in sample B with respect to that of the Fluid sample A.  By adjusting the under-field $S(q)$ profiles of Sample B with Eq.~(4), we observe that the under-field $\sigma$-anisotropy is also strongly reduced (see Table 3). Moreover we can note that the under-field anisotropy of $S(q=0)$, if any, is not detected experimentally and that the value $\sigma_{0}$ = 2.3 nm deduced from the compressibility determination is here close to the value $\sigma$ = 1.9 nm deduced from our under-field model of parts 5 and 6.\\

The cage model developped in parts~5 and 6 is thus very well adapted to an almost "solid" sample as sample B which presents a very low compressibility ($\chi_{T,0}$ = 0.04). However in these glassy conditions, such "freshly prepared" samples \cite{Wanders_PRE09} present with time slow dynamics and ageing properties \cite{Aymeric_EPL06} which are also anisotropic under an applied field \cite{Wanders_BJP09}. The heterogeneous nature of this dynamics has been demonstrated in zero field \cite{Wanders_JPCM08}. One could eventually hypothesize that, in that case, the under-field anisotropy of $\sigma$ transforms in an anisotropy of the heterogeneities. This remains to be studied, in close relation with the local anisotropies evidenced here in the probed concentrated systems.\\

%%%%

\section{Conclusion}

The structure factor $S(\vec q)$ of concentrated aqueous magnetic fluids is here experimentally determined by SAXS and analyzed in the case where the interparticle interactions are repulsive on average. The experiments are performed in controlled conditions of electrostatic repulsion (constant ionic strength) and NP size (using several samples of similar NP diameter). By comparing in zero applied field, the compressibility of the system and the $q_{max}$ value associated to the $S( q)$ maximum we determine (i) the cage dimension (spherical on average), (ii) the mean quadratic displacement of the NPs and (iii) the bulk modulus of the system as a function of the NP volume fraction. Under an applied field, the interparticle interaction remains always repulsive but becomes anisotropic because of the magnetic dipolar interaction contribution and the structure factor then presents anisotropic features on the scale of $2\pi/q_{max}$. The cage becomes anisotropic and presents an oblate deformation at almost constant volume.

To describe these local anisotropic features, we develop a formalism connecting the magnetic and under-field elastic characteristics of the Magnetic NPs system with the values of the scattering vector $q^{max}_{\parallel}=\frac{2\pi}{d_{\parallel}}$ and $q^{max}_{\perp}=\frac{2 \pi}{d_{\perp}}$ at the maxima of $S(\vec q)$, defined respectively in the direction of and normal to the applied field $\vec H$. This formalism is based on the elastic deformation of the cage at constant volume under the applied field. On the scale of the maximum of the structure factor, this model catches the essence of the $q^{max}$ and $S^{max}$ anisotropies observed here and allows to deduce the (Young) elastic modulus of the magnetic fluid associated to its under-field deformation. This Young modulus is of the same order of magnitude as the zero-field Bulk modulus and is larger by a factor of the order of a few units.

However under-field we experimentally observe an anisotropy of the mean quadratic displacement of the NP around their mean position, which is not captured by the model. At very high volume fraction this feature is strongly damped and almost disappears as the sample is becoming glass-forming with a very low compressibility and a large Bulk modulus.\\

\textbf{Acknowledgements}\\

We acknowledge ESRF for the beamtime allocation on ID02 beamline, our local contact P. Panine together with V. Dupuis and T. Narayanan for their help during the experiment.\\

This work has been supported by the international PHC progam OSMOSE n$^{\circ}$ 22497YE, the GDRE GAMAS, the direction of UPMC International Relations and the "F\'ed\'eration 21 de l'UPMC - Dynamique des Syst\`emes Complexes".\\
%%%
%%%%
\textbf{Annex I : Effective field model of the magnetic fluid magnetization $M_{MF}$}\\

To take into account the magnetic interparticle interaction under magnetic field, an effective field model has been developped in the framework of a mean field approximation \cite{Bacri_PRL95, FloGa_PRE02}, to describe the magnetization $M_{_{MF}}$ of concentrated magnetic fluids :
\begin{equation}
M_{_{MF}}=\Phi m_{S}L(\xi_{e})
\label{Eq:AI-1}
\end{equation}
where $\Phi$ is the MF volume fraction, $m_{S}$ the nanoparticle saturation magnetization, $L(\xi_{e})$ = coth($\xi_{e}$)-$\xi_{e}^{-1}$ the Langevin function with $\xi_{e}$ the effective Langevin parameter given by the self-consistent equation :
\begin{equation}
\xi_{e}=\xi +\lambda \gamma L(\xi_{e})
\label{Eq:AI-2}
\end{equation}
with $\xi$ = $\mu_{0}m_{s}H \pi d_{NP}^{3}/6k_{B}T$, $\lambda$ the effective field constant and $\gamma$ the dipolar interaction parameter of the dispersion defined here as $\gamma$ = $\mu_{0}m_{s}^{2}\Phi  \pi d_{NP}^{3} /6k_{B}T$. In $\xi$, we use the average $\sqrt[3]{<d_{NP}^{3}>}$ computed over the whole diameter distribution in each sample. The effective field constant $\lambda$ has been determined to be 0.22 in previous experimental works on similar Magnetic Fluids \cite{Bacri_PRL95, FloGa_PRE02}, as well in numerical simulations\cite{Meriguet_JCP04}. The parameter $\Psi_{dd}=\gamma/\Phi$ is characteristic of the nanoparticles. It is experimentally determined \cite{Guill_MagnHydr12} (see Table 1) by the measurement of the initial susceptibility $\chi_{0}=M/H$ of dispersions at low concentration for which $\gamma=3\chi_{0}$ and $\Psi_{dd}=3\chi_{0}/\Phi$.\\

%%%
%%%%
\textbf{Annex II : Carnahan-Starling Osmotic compressibility}\\

The NP's in this work bear a negative superficial charge which produces a strong electrostatic interparticle repulsion. The Carnahan-Starling formalism \cite{Carnahan_JCP70} is usually used to describe Hard Sphere systems (HS). In the present case the osmotic pressure of the NP's system can be also described in this framework, if effective spheres are introduced in the term correcting the Perfect Gas expression of the osmotic pressure \cite{Cousin_PRE03, Wanders_PRE09}. Introducing the screening length $\kappa^{-1}$ of the NP's system, the volume of these effective spheres is $\frac{\pi}{6}(d_{NP}+2\kappa^{-1})^3$ (instead of $V_{NP}=\frac{\pi}{6}d_{NP}^3$) and their volume fraction is $\Phi_{eff}$ (instead of $\Phi$). The osmotic pressure then is expressed as :
\begin{eqnarray*}
\Pi V_{NP}= k_{B}T \Phi Z_{CS}(\Phi_{eff})   \quad\text{with}  \quad \quad  \quad \quad  \quad \quad \quad \quad \quad \text{}
\end{eqnarray*}
\begin{equation}
\quad  \quad \quad  \quad Z_{CS}(\Phi_{eff})=\frac{1+\Phi_{eff}+\Phi_{eff}^{2}-\Phi_{eff}^{3}}{(1-\Phi_{eff})^{3}}.
\end{equation}
with %For samples A and C at [cit]$_{free}$ = 0.03 M, $\Phi_{eff}$ has been determined in \cite{Wanders_PRE09} by assimilating the HS freezing of the effective spheres ($\sim58\%$) with the experimental freezing volume fraction $\Phi^{*}=25\%$ of the dispersions. It writes  
$\Phi_{eff}\sim \Phi (1+\frac{2\kappa^{-1}}{d_{NP}})^{3}$. The osmotic compressibility, being defined as :
\begin{equation}
\chi_{T,0}=\frac{k_{B}T}{(\partial\Pi V_{NP}/\partial\Phi)_{T}}=\frac{1}{Z_{CS}(1+\frac{\Phi_{eff}}{Z_{CS}}\frac{\partial Z_{CS}}{\partial\Phi_{eff}})}
\end{equation}
can be written as a function of $\Phi_{eff}$ as :
\begin{equation}
\chi_{T,0}=\frac{(1-\Phi_{eff})^{4}}{1+4\Phi_{eff}+4\Phi_{eff}^{2}-4\Phi_{eff}^{3}+\Phi_{eff}^{4}}.
\end{equation}

This expression is compared in Fig.~2-b with the experimental determinations of $\chi_{T,0}$ using $\Phi_{eff}\sim 1.9 \Phi$ and $\kappa^{-1}\sim$ 1.2 nm, close to the evaluations of \cite{Wanders_PRE09}. It fits well the experimental values up to $\Phi\sim20\%$ thus up to $\Phi_{eff}\sim38\%$, close to the customary value for such an effective H.S. model \cite{Vrij_FaradDisc83}.


%\balance


\footnotesize{
%\bibliography{rsc} %your .bib file
%\bibliographystyle{rsc} %the RSC's .bst file
%%%%%

}

\end{document}